\documentclass[referee]{raa}
\usepackage{graphicx,times}
\usepackage{natbib}
\usepackage{amssymb,amsmath}
\bibpunct{(}{)}{;}{a}{}{,}

\begin{document}

   \title{Lijiang 2.4-meter Telescope and its Instruments$^*$
   \footnotetext{\small $*$ Supported by National Natural Science Foundation of China. }
   }

 \volnopage{ {\bf 20XX} Vol.\ {\bf X} No. {\bf XX}, 000--000}

   \setcounter{page}{1}

   \author{
   Chuan-Jun Wang\inst{1,2,3,4},Jin-Ming Bai\inst{1,2,3},Yu-Feng Fan\inst{1,2,3},Jirong Mao\inst{1,2,3},Liang Chang\inst{1,2,3,4},
   Yu-Xin Xin\inst{1,2,3,4},Ju-Jia Zhang\inst{1,2,3},Bao-Li Lun\inst{1,2,3},Jian-Guo Wang\inst{1,2,3},Xi-Liang Zhang\inst{1,2,3},
   Mei Ying\inst{5},Kai-Xing Lu\inst{1,2,3},Xiao-Li Wang\inst{1,2,3},Kai-Fan Ji\inst{1,3}, Ding-Rong Xiong\inst{1,2,3},
   Xiao-Guang Yu\inst{1,2,3},Xu Ding\inst{1,2,3},Kai Ye\inst{1,2,3},Li-Feng Xing\inst{1,2,3},Wei-Min Yi\inst{1,2,3}, Liang Xu\inst{1,2,3}, 
   Xiang-Ming Zheng\inst{1,2,3},Yuan-Jie Feng\inst{1,2,3},Shou-Sheng He\inst{1,2,3},Xue-Li Wang\inst{1},Zhong Liu\inst{1},
   Dong Chen\inst{1}, Jun Xu\inst{1},Song-Nian Qin\inst{1},Rui-Long Zhang\inst{1},Hui-Song Tan\inst{1},Zhi Li\inst{1},
   Ke Lou\inst{1},Jian Li\inst{1}, Wei-Wei Liu\inst{1}
   }
   \institute{ Yunnan Observatories, Chinese Academy of Sciences, 396 Yangfangwang, Guandu District, Kunming 650216, P.R. China; {\it wcj@ynao.ac.cn};{\it baijinming@ynao.ac.cn}\\
    \and
             Key Laboratory for the Structure and Evolution of Celestial Objects,Chinese Academy of Sciences, 396 Yangfangwang, Guandu District, Kunming 650216, P.R. China\\
	\and
             Center for Astronomical Mega-Science, Chinese Academy of Sciences, 20A Datun Road, Chaoyang District, Beijing, 100012, P.R. China \\
    \and
             University of Chinese Academy of Sciences,Beijing 100049, P.R. China\\
      \and 
             Astrophysics Center of Guangzhou University, Guangzhou, 510006, P.R. China\\
\vs \no
   {\small Received 20XX Month Day; accepted 20XX Month Day}
}

\abstract{Lijiang 2.4-meter Telescope(LJT), the largest common-purpose optical telescope in China, has been applied to the world-wide astronomers since 2008. It is located at Gaomeigu site, Lijiang Observatory(LJO), the southwest of China. The site has very good observational conditions. Since 10-year operation, several instruments have been equipped on the LJT. Astronomers can perform both photometric and spectral observations. The main scientific goals of LJT include photometric and spectral evolution of supernova, reverberation mapping of active galactic nucleus, physical properties of binary star and near-earth object(comet and asteroid), identification of exoplanet, and all kinds of transients. Until now, the masses of 41 high accretion rate black holes have been measured, and more than 168 supernova have been identified by the LJT. More than 190 papers related to the LJT have been published.
In this paper, the general observation condition of the Gaomeigu site is introduced at first. Then, the LJT structure is described in detail, including the optical, mechanical, motion and control system. The specification of all the instruments, and some detailed parameters of the YFOSC is also presented. Finally, some important scientific results and future expectations are summarized.
\\
}

   \authorrunning{C.-J. Wang et al. }            
   \titlerunning{Lijiang 2.4m telescope and its instruments}  
   \maketitle

%
\section{Introduction}           
\label{sect:intro}

The Lijiang 2.4-meter Telescope belongs to Lijiang Observatory(IAU code O44), Yunnan Observatories (YNAO), Chinese Academy of Science (CAS). The location of the observatory is  $100^{\circ}1^{\prime}$48"(E), $26^{\circ}41^{\prime}$42"(N), with the altitude of 3193 meters (Figure 1). The observational conditions of the site are summarized  in Table 1 \citep{2017PA...35...367}. The site survey for the 2.4-meter telescope began in 1994, and the construction of the Lijiang 2.4-meter telescope as well as the Lijiang Observatory began in 2003. The telescope was under testing with instruments installation from 2008, and it has been applied to astronomers for astronomical observations since 2012.

\begin{table}
\bc
\begin{minipage}[]{100mm}
\caption[]{Observation Conditions of Lijiang Observatory\label{tab1}}\end{minipage}
\setlength{\tabcolsep}{10pt}
\small
 \begin{tabular}{lcc}
  \hline\noalign{\smallskip}
  Average Observation Time(hours) & 2150(2012 to 2014) & 2000(2015 to 2017) \\
  Sky Brightness & V:22.06 mag/sq.arcsec & 22.14 mag/sq.arcsec \\
  Extinction Coefficient & KV=0.14 & KB=0.3 \\
  Seeing(Average) & 0.97"(2014) & 1.0"(2015) \\
  \noalign{\smallskip}\hline
\end{tabular}
\ec
\tablecomments{0.86\textwidth}{We utilize the Sky Quality Meter (SQM-LE) produced by Unihedron Company to get the night-time sky brightness. }
\end{table}

\begin{figure}
	\centering
	\includegraphics[angle=0]{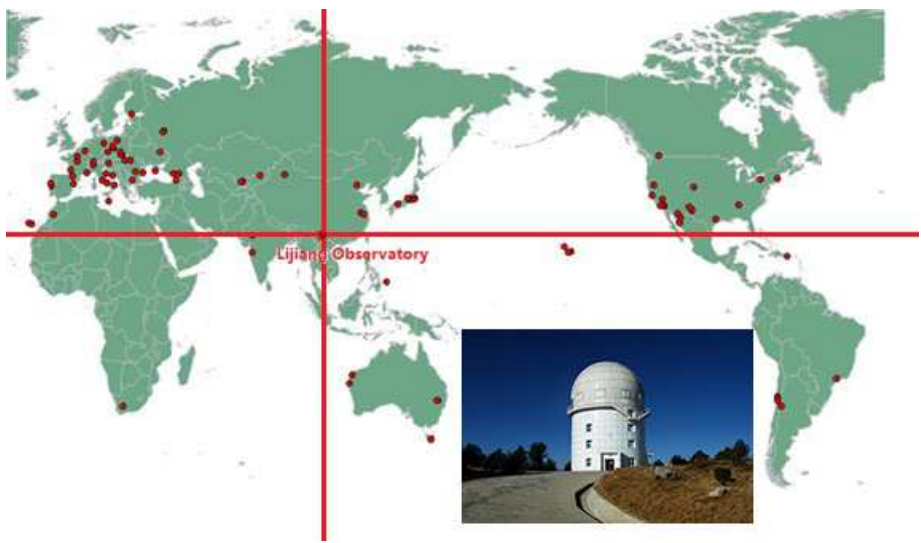}
	\caption{Lijiang Observatory Location and 2.4-meter Telescope Dome}
	\label{fig1}
\end{figure}

The Lijiang 2.4-meter telescope is the largest optical telescope at present in China for general astronomical observation. It can play more and more important roles in the future. The telescope was designed by the Telescope Technology Limited(TTL) company in Liverpool, England, and it provides very stable and precise performance. This telescope has a traditional altitude-azimuth reflecting type that comprises a Ritchey-Chretien Cassegrain optical design. The main body of the telescope is under truss structure and supported by the hydrostatic bearing system. The primary mirror is supported both axially and laterally by a three-sector pneumatic support system. The telescope incorporates a real-time motion control system to have accurate position in altitude, azimuth and Cassegrain axes of rotation. Additionally, the auto-guider selects star image centroids in the field-of-view to provide refine tracking corrections to the control system. The pointing accuracy of this telescope is better than 2 arcsecs, the open-loop tracking accuracy is better than 2 arcsecs/hour, and the closed-loop tracking accuracy is better than 0.5 arcsecs/hour. The telescope and the dome can be controlled automatically with embedded industrial PCs and real-time operating system, that provides a possibility for remote and autonomous observation. The structure of the telescope is shown in Figure 2.

Although the telescope is small compared to other large telescopes in the world, many kinds of scientific investigations can still be carried out due to the important geographical location of the site. It can take photometric observations with standard Johnson and SDSS filters. Both low/medium and high-resolution spectral observations can be also performed. The main scientific goals of the telescope focus on time-domain astronomy, including photometric and spectroscopic evolution of supernova, reverberation mapping of active galactic nucleus, physical properties of binary star and near-earth object (comet and asteroid), identification of exoplanet, gamma-ray burst follow-up, and gravitational wave electromagnetic counterpart observation. In addition, some other scientific issues, such as high-redshift quasars, Li-rich stars, and precise CCD position of natural satellites of planets in the solar system, are also included. The instruments mounted on the telescope are Yunnan Faint Object Spectrograph and Camera (YFOSC), Li-Jiang Exoplanet Tracker (LiJET), High Resolution Spectrograph (HiRES), Multicolor Photometric System(PI CCD) and China Lijiang IFU(CHILI). Thanks to the Rapid Instrument Exchanging System that we installed in 2012, observers can switch to a certain instrument quickly and can perform the semi-simultaneous photometric and spectral observation for the same target. Thus, the telescope can perform several kinds of observations at a single night \citep{2015RAA....15..918F}.  

We describe the structure of the 2.4-meter telescope in detail, including optical, mechanical, motion and control systems, in section 2. The instruments of the telescope are introduced in section 3. Some important equipment related to the telescope are presented in section 4. Some observational outputs of the telescope are summarized in section 5. Some expected progresses in the future are written in section 6.

\begin{figure}
	\centering
	\includegraphics[width=12.0cm, angle=0]{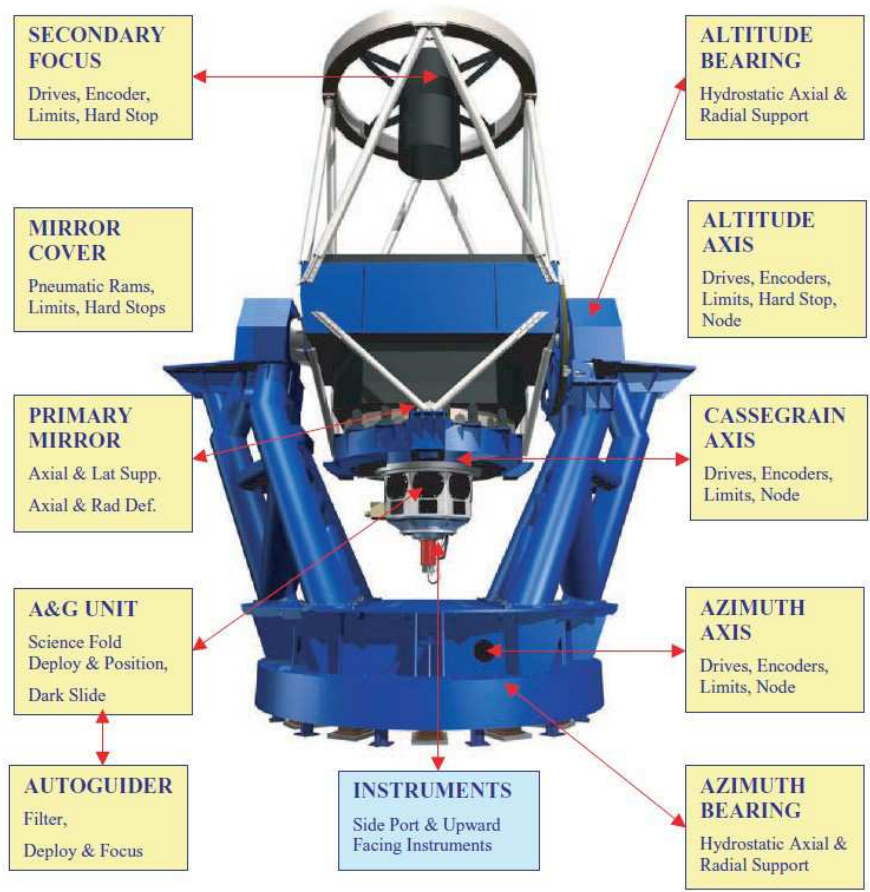}
	\caption{The 2.4-meter Telescope System }
	\label{fig2}
\end{figure}

\section{The Structure of 2.4-meter Telescope}
\label{sect:Telescope}
The 2.4-meter telescope is comprised by four main subsystems, they are optical system, mechanical system, motion system and control system. All of them are described in detail in the following subsections.

\subsection{Optical System}
The optical design of the primary and secondary mirrors is Ritchey$-$Chretien Cassegrain structure, which provides one Cassegrain focus and two Nasmyth focus. The on-axis field-of-view of cassegrain focus is 10 arc minutes, and it can reach 40 arc minutes by using correction mirror. By using the $45^{\circ}$ folding mirror installed in the A\&G box at the cassegrain, we can obtain one straight port and eight side ports at the cassegrain focal station. Thus, we can install nine instruments on the telescope at the same time. The nasmyth focal station can be switched by setting the third mirror into the tube that mounted from the telescope center section. The nasmyth focal station can provide a field-of-view of 8 arc minutes. The telescope image is kept in focus by moving the secondary mirror axially. The optical specification of the telescope is listed in Table 2, and the optical diagram of the telescope is illustrated in Figure 3 (also see \citealt{2015RAA....15..918F}).

\begin{figure}
	\centering
	\includegraphics[width=14.0cm,angle=0]{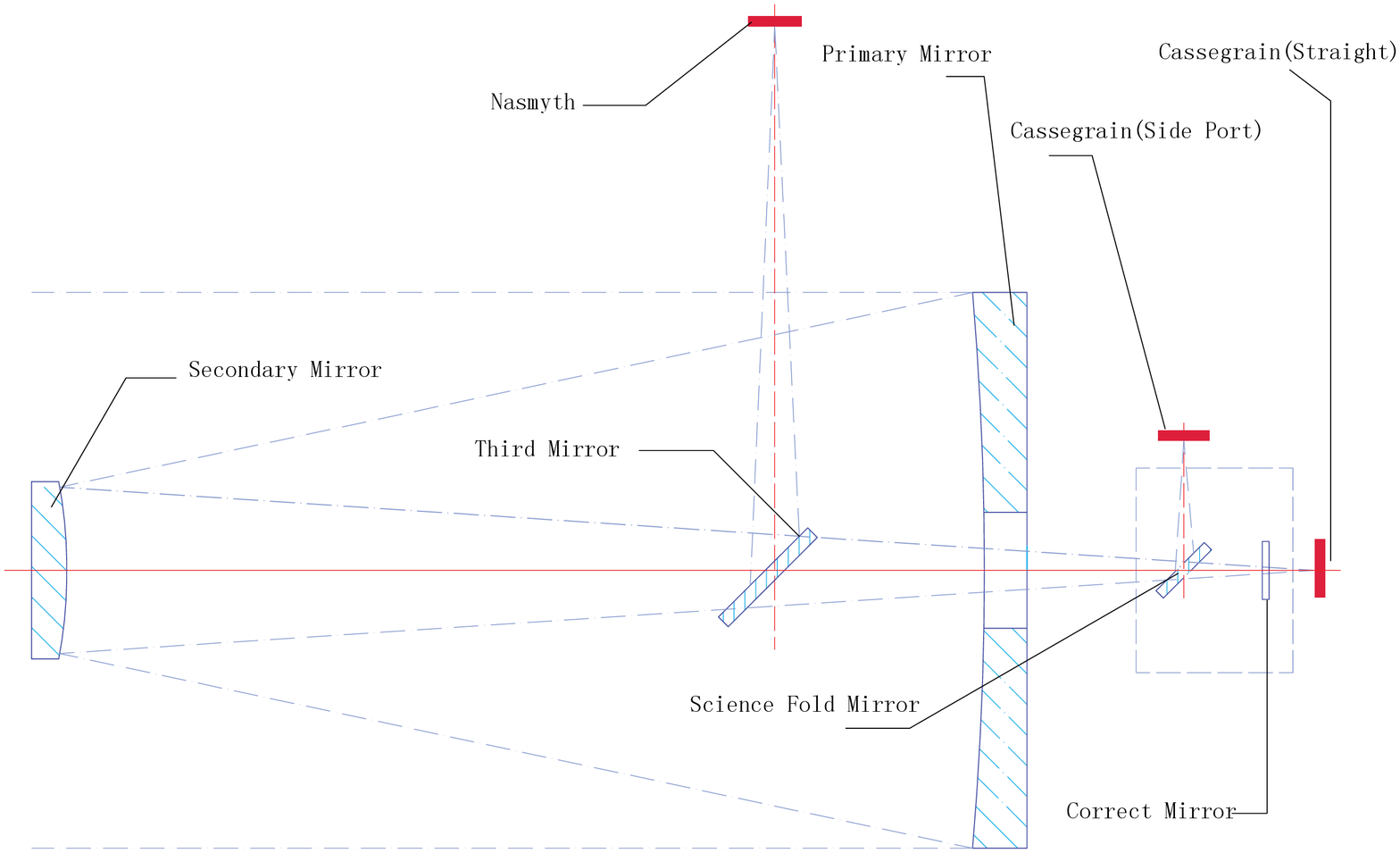}
	\caption{Optical Diagram for the 2.4-meter Telescope}
	\label{fig3}
\end{figure}

\begin{table}
\bc
\begin{minipage}[]{140mm}
\caption[]{Telescope Optical Specification\label{tab2}}\end{minipage}
\setlength{\tabcolsep}{10pt}
\small
 \begin{tabular}{ccc}
  \hline\noalign{\smallskip}
   & Clear Aperture & 2400mm \\
   & Central Bore & 500mm \\
  Primary Mirror & Focal Ratio & F/2.43 \\
   & Radius of Curvature & -11520mm \\
   & Conic Constant & -1.073 \\
  \hline\noalign{\smallskip}
   & Clear Aperture & 709mm \\
   & Radius of Curvature & -4760.44mm \\
  Secondary Mirror & Conic Constant & -4.187 \\
   & Distance to Primary Mirror & 4094.114mm \\
   & Distance to Focal Plane & 5550.870mm \\
  \hline\noalign{\smallskip}
   & Focal Ratio & F/8 \\
  Cassegrain focus & FOV of Fold Port & 8 arc minutes \\
   & FOV of Straight Port & 10 arc minutes \\
   & Corrected FOV of Straight Port & 40 arc minutes \\
  \hline\noalign{\smallskip}
  Nasmyth focus & Focal Ratio & F/8 \\
   & FOV & 8 arc minutes \\
   \hline\noalign{\smallskip}
\end{tabular}
\ec
\end{table}

The concave primary mirror is made of Zerodur, a glass-ceramic material with a nearly zero thermal coefficient of expansion. It is coated with Aluminum. The back of the primary mirror is flat. This is convenient for the axial support pads bear onto it during operation but not physically attached. Invar pads are bonded to the outer edge of the mirror for attaching both the lateral support system and the mirror defining system. The primary mirror is shown in the left of Figure 4. The convex secondary mirror is also made of Zerodur. The back of it is flat with a central bore. Invar pads are bonded to the back surface of the mirror for attaching the axial support system, while the lateral support is bonded inside the rear borehole. The secondary mirror is shown in the right of Figure 4.

\begin{figure}
	\centering
	\includegraphics[width=14.0cm, angle=0]{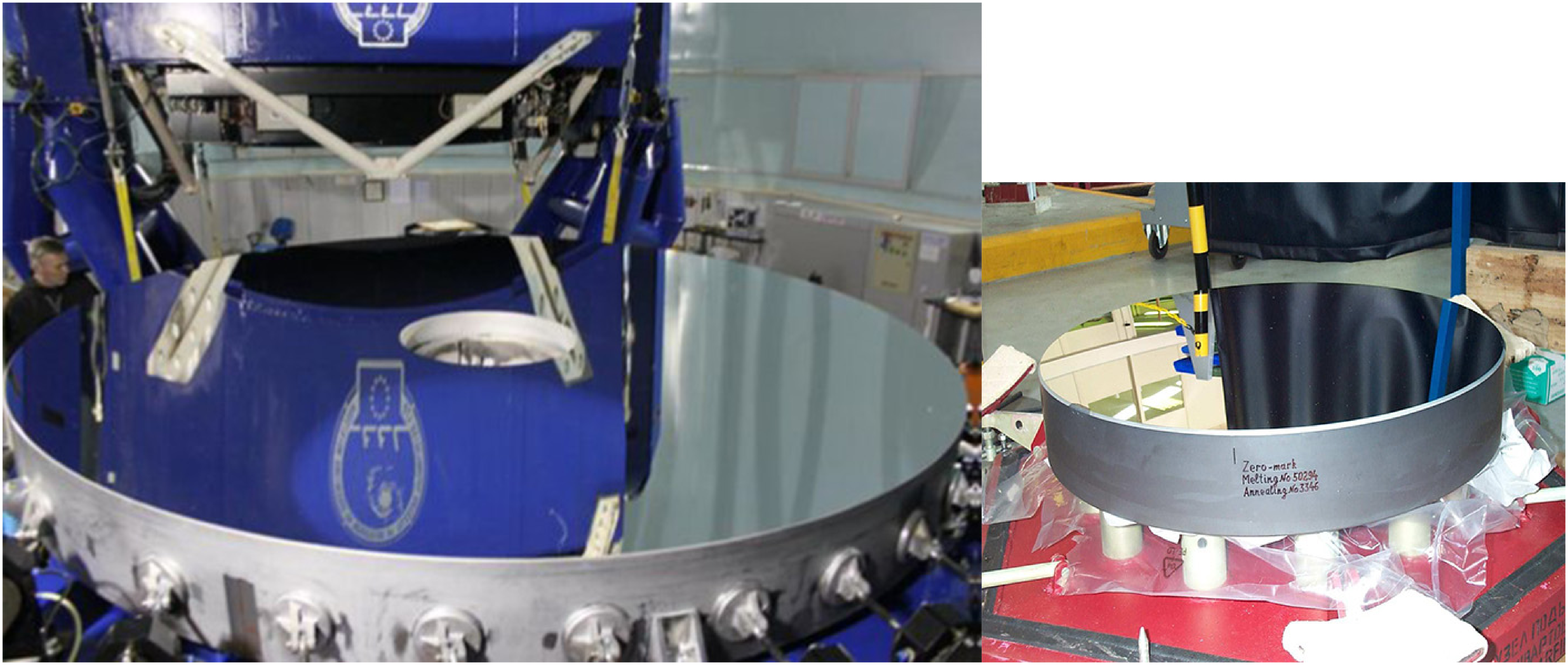}
	\caption{The primary and secondary mirror}
	\label{fig4}
\end{figure}

There are two light baffles mounted on the telescope structure to prevent stray light from entering the focus of the telescope.The upper baffle is mounted from the secondary mirror assembly, and the lower baffle is mounted from the telescope center section. They are shown in Figure 5.  The central beam obscuration is mainly caused by the secondary assembly. The supporting vanes of the secondary mirror and the lower baffle induce a small amount of additional obscuration. The total obscuration is 16.67\%.

\subsection{Mechanic System}
The main body of the telescope is fabricated from steel plate and steel tube, which are welded to have high stiffness and  low mass structure. The azimuth and altitude axes are supported by the hydrostatic bearings. The azimuth axis takes three load bearings and three guide pads, and the altitude axis is supported by four load bearings and two guide bearings.

The telescope tube is the structure that carries the principal optical components of the telescope. The optical axis of the tube, defined by the Cassegrain rotation bearing, is arranged to intersect and be orthogonal with the altitude axis of rotation. The center section is the foundation of the telescope tube, which has two trunnion bearings attached at the two opposite sides using hydrostatic bearing pads to support the entire telescope tube. The tube is a traditional open-truss Serrurier construction that can maintain the optical alignment. The lower tube structure connects the center section with the primary mirror and the Cassegrain assembly. The upper tube structure is connected to the secondary mirror assembly. The mirror cover is mounted from the upper surface of the center section and it can be controlled by the high-pressure air. The lower baffle is also mounted from the center section.  We can mount a tertiary mirror with 45 degree into the main light path of the telescope to direct the light to the Nasmyth focal station.

The primary mirror cell is incorporated with the primary support system, one Cassegrain rotator bearing and one Cassegrain rotator cable wrap. The structure of the primary cell is designed with a stiff-under structure to minimize distortions between the truss locations and the Cassegrain rotator bearing. The 2.3-tone primary mirror is supported by pneumatic actuators both axially and laterally. The axial units are divided into three sectors. Each sector has an axial-defining unit situated in the middle of it, and mounted at the outside edge of the mirror. A load cell in these units provides force feedback information for the mirror support system, which can be used to correct any detected load by adjusting the air pressure in the axial actuators of the appropriate sector.

The top-end ring incorporates the interface ring and the upper telescope tube. It is used to support the secondary mirror cell. It is  connected to the center section by the top interface. The secondary mirror cell can be disconnected from the interface ring without disturbing the alignment of the telescope tube.

The mechanic structure of the telescope tube is shown in Figure 5.

\begin{figure}
	\centering
	\includegraphics[width=10.0cm, angle=0]{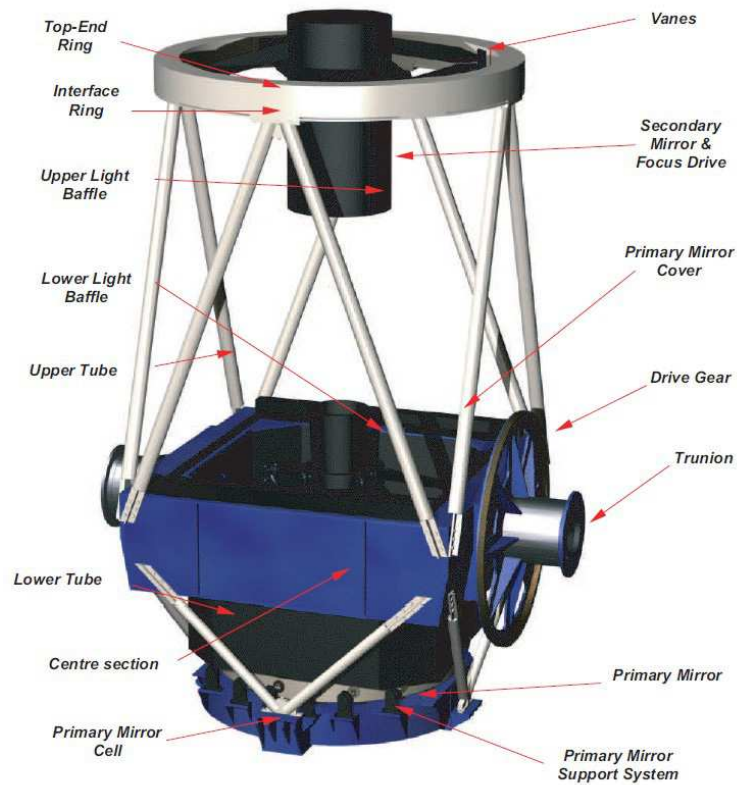}
	\caption{The mechanic structure of the telescope tube}
	\label{fig5}
\end{figure}
	
\subsection{Motion System}
The motion system of the telescope incorporates all the motive elements, and is controlled by the Master Control System(MCS) from inputs supplied by the Telescope Control System(TCS) or the Engineering Control Interface(ECI). The motive elements include: azimuth, altitude, rotator, primary mirror support, mirror cover, secondary mirror focus assembly, auto-guider and science fold mirror. The main functions of motion system are: \\
- Move the azimuth and altitude. \\
- Drive the secondary mirror focus. \\
- Drive the Cassegrain rotator. \\
- Support the primary mirror and drive the mirror cover. \\
- Deploy the auto-guider, move the auto-guider focus and deploy the filter. \\
- Deploy the science fold mirror and rotator it to the specified side port. \\

Both azimuth and altitude are driven by a pair of electric motors in the anti-backlash configuration. Each motor is connected to a gearbox and the outputs are used to drive a pinion. The position information is provided by a Heidenhain optical tape encoder adhered inside of the bearing, which can be transferred to the Telescope Control System.

In order to compensate the effect that the image on the focal plane rotates with the Azimuth motion of the telescope when the telescope tracks a scientific target in Azimuth and Altitude, the Cassegrain rotator is used to de-rotate the image. The Cassegrain rotator is also driven by a pair of electric motors in an anti-backlash configuration, and the position information is also provided by the Heidenhain optical tape encoder.

\subsection{Control System}
The control system of the telescope can be divided into Safety Interlock System, Services System, and Computing System.

Safety Interlock System: Independent Programmable Logic Controllers(PLC) are used to manage subordinate and supporting services for the telescope and the enclosure. This system includes the control of interlocks between local and remote control. The control limit and the emergency-stop(E-stop) can prevent from major damage to the telescope structure, electrical, pneumatic and optical system. The whole telescope system will be shut down if one of the emergency-stop is triggered at anywhere in the dome to protect the staffs and the science instruments.

Service System: Service system includes enclosure PLC and services PLC. The Enclosure PLC(EPLC) mainly handle opening and closing the aperture, as well as the tracking of the target along with the telescope. The Service PLC(SPLC) mainly handle all the auxiliary actions and services associated with the operation of the telescope. These actions include: electrical power control, electrical power distribution, local/remote control, control and monitoring of the hydrostatic bearing system, control and monitoring of the primary mirror support system, opening and closing primary mirror cover, control and monitoring of the motion axis, control the telescope manually, on/off mount interlocking system, temperature monitoring, alarm handling, temperature monitoring of cooling system, and local control.

Computing System: Computing system consists of several on-mount and off-mount industrial computers. The on-mount computers are Azimuth node, Altitude node, Cassegrain node, auxiliary mechanism node(AMN) and auto-guider control computer(ACC). The off-mount computers are Supervisory Control Computer(SCC, controls the start-up, shutdown and monitoring of all computing systems), Master Control Computer(MCC, to run the database that including all telescope parameters, to run the engineering control interface and remote control, and monitoring the enclosure), and Telescope Control Computer(TCC, to run the Telescope Control System that can receive the requirements or commands from the operator, and to perform astrometric transformation and send it back to MCS to perform the properly operation of the telescope). All on-mount and off-mount computers are connected to the telescope local area network, and all computers apart from the TCC are using the real-time operating system QNX. QNX is a network-wide operating system and provides an inherent network file system. Therefore, all on-mount computers are diskless nodes that can net-boot from the MCC during the startup period. After startup, all the on-mount computers apply the file system of MCC and SCC through the network.

\section{INSTRUMENTATION INCORPORATED}
\label{sect:Instrument}
There are five scientific instruments mounted on the Cassegrain focal station. They are YFOSC, Multicolor Photometric System(PI CCD), HiRES, LiJET and CHILI. All of these instruments are described in detail in the following subsections.

\subsection{YFOSC}
YFOSC is the most popular instrument on the 2.4-meter telescope that is installed on the straight through port of Cassegrain. It is a scientific instrument for multi-mode observation based on focal reducer. It can perform photometry and low/medium dispersion spectal observations semi-synchronously. Time-domain observation is one of the major aim for the telescope, and we usually arrange different kinds of observations in one night. Thus, we need switch observational modes frequently in short time. We can easily use YFOSC to achieve different scientific aims by quickly switching observational modes. During the night, one observer can attempt different targets and perform different observations depending on the different observing conditions. It is convenient to change among different filters, grisms and slits. Its structure is shown in figure 6. It has five wheels: the aperture wheel, the YFUA and YFUB wheels, the filter wheel, and the grism wheel. All the wheels are controlled by the integrated control system to change among different elements and to perform different observations.

The CCD chip for YFOSC is a back illuminated type CCD chip produced by e2V company. The CCD controller uses all-digital hyper-sampling technology called $3^{rd}$ generation CCDs(CCD3), which can keep low readout noise under high readout speed. The parameters are listed in table 3.All the optical elements installed onto YFOSC (except filters) are listed in table 4.

For photometric observation, the light goes through the collimating mirror, filters, imaging mirror. Then, the image is on the CCD chip under a focus ratio conversion from F/8 to F4.1. This conversion can reduce the affect produced by the oversampling of the CCD and improve the ultimate detection capability. Using SDSS r-band filter with the exposure time of 20 minutes, the limiting magnitude of point source is about 23.5mag with the S/N ratio $\geq$ 3 \citep{2014ART...11...176}. Observers can reduce the readout section and use binning to reduce the readout time if it is necessary. When suitable aperture and grism are moved into the light path, YFOSC can change to spectral observing mode. YFOSC can perform long-slit spectrum and cross-dispersion spectrum. Both kinds of spectra can cover the wave-length from 340nm to 980nm. The long-slit spectrum mode with single grism can get low-resolution spectrum. Using the long-slit spectrum with exposure time of 30 minutes, the limiting magnitude is 19.5mag by 1.8" aperture and the resolution is 300/pix with S/N ratio about 10. The target to get this result is  SN2011fe \citep{2016ApJ...820...67Z}.The cross-dispersion spectrum mode uses low dispersion grism and medium dispersion echelle to get medium resolution spectrum. The spectral resolution is 10000/pix under the magnitude limit of 15.5mag by 0.85" aperture \citep{2018MNRAS.481...878Z}. The detailed spectral efficiency of each grism using long slit was given by \cite{2012ART...9...411F}.

\begin{table}
\bc
\begin{minipage}[]{140mm}
\caption[]{Parameters of CCD for YFOSC\label{tab3}}\end{minipage}
\setlength{\tabcolsep}{10pt}
\small
 \begin{tabular}{cc}
  \hline\noalign{\smallskip}
  Parameter & Value \\
  \hline\noalign{\smallskip}
  Pixels & 2048\x4608 \\
  Pixel Size & 13.5um\x13.5um \\
  Image Area & 27.6mm\x62.2mm \\
  Field of View & 9.60'\x9.60' \\
  (Photometry) & (2K\x2K) \\
  Image Scale & 0.283"/pixel \\
  Cooling Mode & Liquid Nitrogen: -$120^{\circ}$C \\
  Gain & 0.33e- \\
  Readout Noise & 6.3e-(Speed: 400kpixels/s) \\
   & $<$5.0e-(Speed: 200kpixels/s) \\
  \noalign{\smallskip}\hline
\end{tabular}
\ec
\end{table}

\begin{figure}
	\centering
	\includegraphics[width=10.0cm, angle=0]{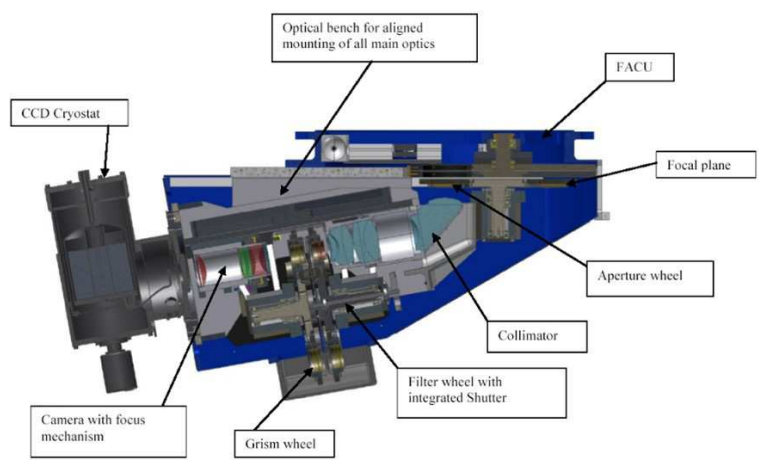}
	\caption{The  structure of YFOSC}
	\label{fig6}
\end{figure}

\begin{table}
\bc
\begin{minipage}[]{140mm}
\caption[]{Parameters of major optical elements for YFOSC\label{tab4}}\end{minipage}
\setlength{\tabcolsep}{10pt}
\small
 \begin{tabular}{cccccccc}
  \hline\noalign{\smallskip}
   & & &  Size(um) & & Sky angle(") & &\\
   & & &  54 & & 0.58 & &\\
   & & &   74 & & 0.8 & &\\
   & & &   93 & & 1.0 & &\\
   & & &   112 & &1.2 & &\\
   & Long Slit & & 140 & & 1.5  & & \\
   & & &   168 & & 1.8 & &\\
   & & &  233 & & 2.5 & & \\
   & & &  470 & & 5.0 & &\\
   & & &  940 & & 10.0 & &\\
  \hline\noalign{\smallskip}
   & & & 54 \x 500 & & 0.58 \x 5.37 & &\\
   & & & 74 \x 500 & & 0.8 \x 5.37 \\
   & Short Slit  & & 100 \x 500 & & 1.07 \x 5.37 \\
   & & & 140 \x 500 & &  1.5 \x 5.37 \\
   & & & 460 \x 500 & & 4.94 \x 5.37 \\
   & & & 940 \x 500 & & 10.0 \x 5.37 \\
  \hline\noalign{\smallskip}
  Grism & $\lambda$c & $\lambda$Blaze & Grooves & Dispersion & Resolution & Sp. Range & Order  \\
  No. & (nm) & (nm) & (nm/mm) & (nm/pix) & (@600nm/pix) & (nm) & Range \\
  12 & 730 & 700 & 75 & 1.1 & 545 & 520-980 & 1 \\
  10 & 380 & 390 & 150 & 0.79 & 760 & 340-900 & 1 \\
  3 & 390 & 430 & 400 & 0.29 & 2068 & 340-910 & 1 \\
  15 & 586 & 527 & 300 & 0.39 & 1540 & 410-980 & 1 \\
  5 & 650 & 700 & 300 & 0.46 & 1300 & 496-980 & 1 \\
  14 & 463 & 428 & 600 & 0.17 & 3520 & 360-746 & 1 \\
  8 & 650 & 700 & 600 & 0.15 & 4000 & 510-960 & 1 \\
  13 & & & 316 & 0.06 & 10000 & 340-980 & 3,4,5 \\
  9 & & & 79 & 0.06 & 10000 & 340-980 & 7-23 \\
   \noalign{\smallskip}\hline
\end{tabular}
\ec
\tablecomments{0.86\textwidth}{We utilize the standard Johnson and SDSS filters for photometric observation, and the parameters of these filters are not listed in this table.}
\end{table}

\subsection{Multicolor Photometric System}
Multicolor photometric system consists a PI VersArray 1300B CCD (PI CCD) camera and a set of standard Johnson/SDSS filters. PI CCD is a full-frame and back illuminated CCD produced by the Princeton Instruments company. It is the first scientific instrument used on Lijiang 2.4-meter telescope, and can be used for the common photometric and spectral observations. The major parameters of this CCD camera are listed in table 5. The image scale of PI CCD is  0.20"/pixel, which can match the photometric requirement of the telescope. The high quantum efficiency and the low readout noise make it more suitable for photometric observation. Nowadays the multicolor photometric system is installed at one of the side ports as a backup photometric camera.

\begin{table}
\bc
\begin{minipage}[]{140mm}
\caption[]{Parameters of PI CCD\label{tab5}}\end{minipage}
\setlength{\tabcolsep}{10pt}
\small
 \begin{tabular}{cc}
  \hline\noalign{\smallskip}
  Parameter & Value \\
  \hline\noalign{\smallskip}
  Pixels & 1300\x1340 \\
  Pixel Size & 20um\x20um \\
  Image Area & 26.0mm\x26.8mm \\
  Field of View & 4.40'\x4.48' \\
  Cooling Mode & Liquid Nitrogen: \\
   & -$70^{\circ}$C to -$110^{\circ}$C, $+$/$-$ $0.05^{\circ}$C \\
  Linearity & $<$ 1\%(100kHz), $<$2\%(1MHz) \\
  Readout Noise & 2.84e-(Low speed, Low noise mode) \\
   & 16.3e-(High speed, High gain mode) \\
  \noalign{\smallskip}\hline
\end{tabular}
\ec
\end{table}

\subsection{HiRES}
HiRES is a high-resolution fiber spectrograph developed by Nanjing Institute of Astronomical Optical \& Technology and Yunnan Observatories. It has been put into operation since November, 2015. The optical diagram of the spectrograph is shown in Figure 7. We can use HiRES to perform high-resolution spectral observation.

Most of the optical elements of HiRES are mounted on an independent optical platform placed in the spectral room at the grand floor outside of the dome. The spectral room can maintain very stable temperature($28\pm0.25^{\circ}C$) and pressure($30\pm1$Pa), in order to match the requirements of high-precision radial velocity observation. There have two fibers installed on the Cassegrain of the telescope. One fiber with the diameter of 2.0" provides the spectral resolution of 32,000 at 550nm. The other with the diameter of 1.2" provides the spectral resolution of 49,000 at 550nm. The wavelength coverage of HiRES is from 320nm to 920nm. Observers can choose the suitable fiber depending on the observation requirement and the weather condition. There also has a photomultiplier tuber installed in HiRES to improve the pointing and focusing of telescope and to estimate the exposure time for the target. The CCD chip of HiRES is also produced by e2V company. The chip is also back illuminated type and it is more sensitive in the blue-end. Detailed parameters of this CCD are listed in table 6. The HiRES uses Th-Ar lamp for wavelength calibration at present, and we will add iodine vapor as one reference to improve the accuracy of the calibration in the future. Limited by the star monitoring system, the observational limit magnitude to use HiRES is 13mag under the resolution of 32,000 with 1800 seconds exposure at 550nm with the S/N ratio about 10. The efficiency curve of the whole system is shown in figure 8. The relation between the s/n and the exposure time for different magnitude at 550nm with different fibers is shown in figure 9.

\begin{table}
\bc
\begin{minipage}[]{140mm}
\caption[]{Parameters of CCD for HiRES\label{tab6}}\end{minipage}
\setlength{\tabcolsep}{10pt}
\small
 \begin{tabular}{cc}
  \hline\noalign{\smallskip}
  Parameter & Value \\
  \hline\noalign{\smallskip}
  Pixels & 4096\x4096 \\
  Pixel Size & 12um\x12um \\
  Image Area & 49.2mm\x49.2mm \\
  Cooling Mode & TEC semiconductor cooling: $~$-$90^{\circ}$C \\
   & (With water cycle cooling) \\
  Readout Noise & $<$5.0e-(Readout speed: 50kHz) \\
   & $<$7.0e-(Readout speed: 250kHz) \\
  \noalign{\smallskip}\hline
\end{tabular}
\ec
\end{table}

\begin{figure}
	\centering
	\includegraphics[width=14.0cm, angle=0]{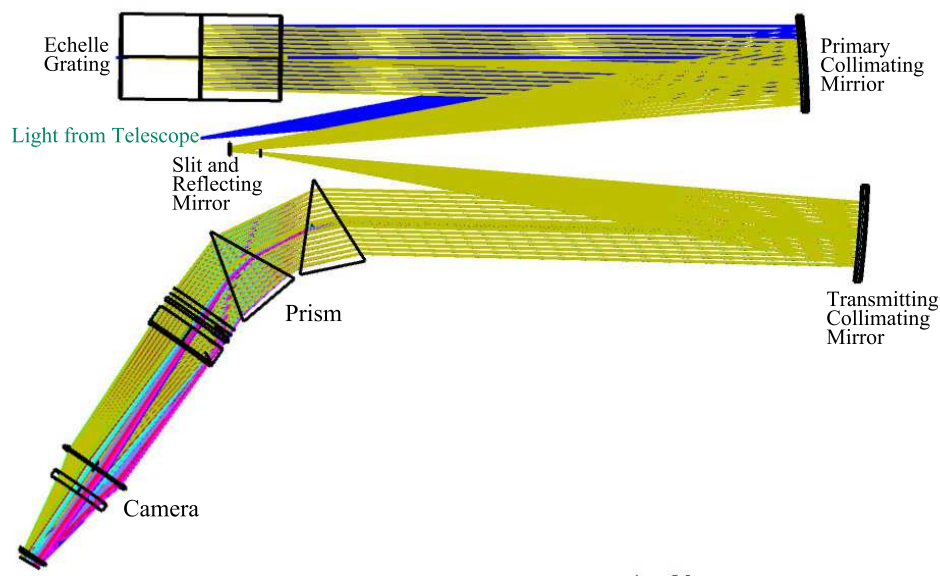}
	\caption{The light path of HiRES}
	\label{fig7}
\end{figure}

\begin{figure}
	\centering
	\includegraphics[width=14.0cm, angle=0]{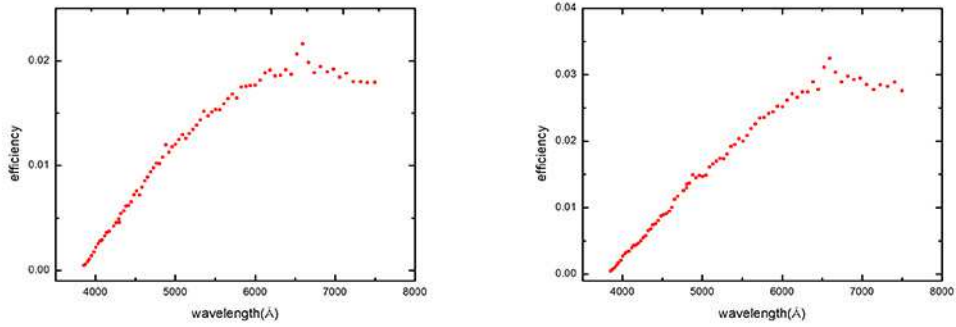}
	\caption{Efficiency curve of HiRES with different fibers (left: fiber diameter of 1.2", right: fiber diameter of 2.0")}
	\label{fig8}
\end{figure}
\begin{figure}
	\centering
	\includegraphics[width=14.0cm, angle=0]{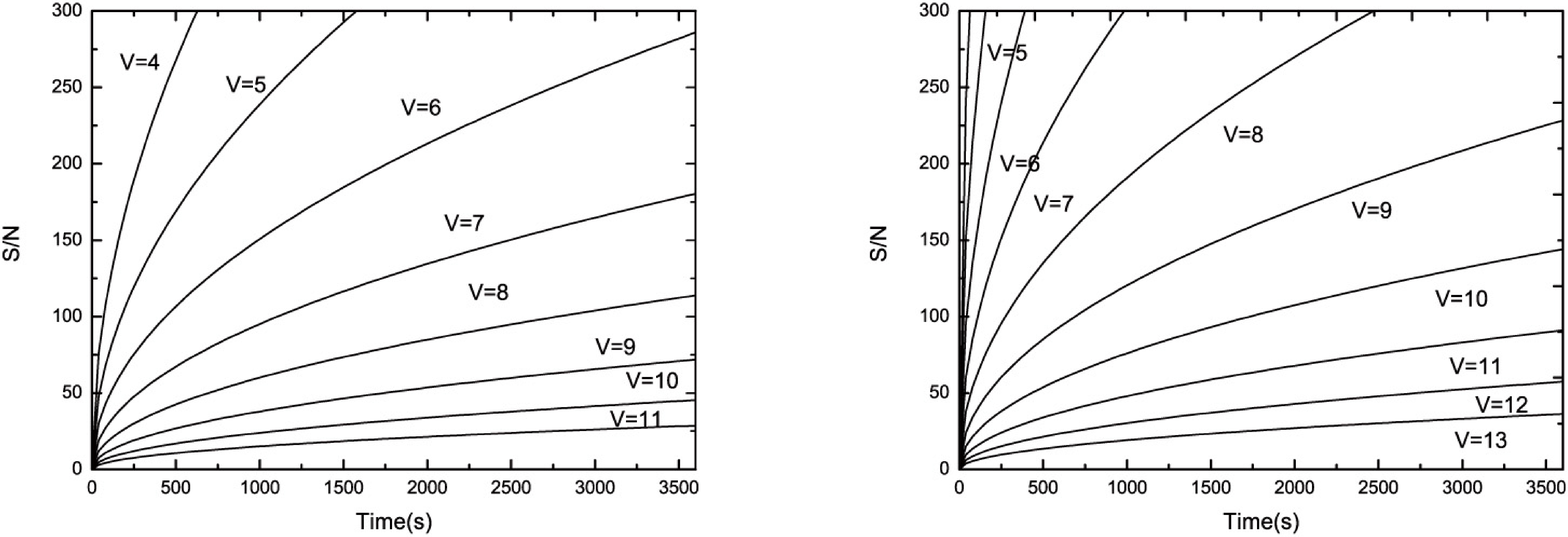}
	\caption{The relation between the S/N and the exposure time for different magnitude at 550nm with different fibers (left: fiber diameter of 1.2", right: fiber diameter of 2.0")}
	\label{fig9}
\end{figure}

\subsection{LiJET}
LiJET is a fiber spectrograph that can provide high-precision radial velocity measurements. This instrument is especially for the exoplanet observation, and it can be also useful for other high-resolution spectral observations \citep{2010AAS...21534803G}. It has two observational modes. One is DEM (Direct Echelle Mode) mode, which has a resolution of 30,000 with the 1.0" slit, and it covers the wavelength between 390nm and 1,000nm. Another is DFDI (Dispersed Fixed-Delay Interferometry) mode or Radial Velocity Mode(RVM), which has a resolution of 18,000 with the 1.6" slit, and it covers the wavelength between 390nm and 690nm with the spectral order from 29 to 52. Under DFDI mode, the fixed delay interferometer comprised with the medium-precision spectrograph to measure the stellar Doppler effect for the exoplanet detection. The precision of the radial velocity calibration is 2.5m/s under the DFDI mode. The optical structure of these two modes are show in figure 10 and figure 11. In order to detect the high-precision radial velocity, LiJET has high-precision temperature and pressure control. The temperature and pressure are very stable, which can achieve 0.1\% precision of the variability. The parameters of the LiJET CCD are listed in table 7.

\begin{table}
\bc
\begin{minipage}[]{140mm}
\caption[]{Parameters of CCD for LiJET\label{tab7}}\end{minipage}
\setlength{\tabcolsep}{10pt}
\small
 \begin{tabular}{cc}
  \hline\noalign{\smallskip}
  Parameter & Value \\
  \hline\noalign{\smallskip}
  Pixels & 4096\x4096 \\
  Pixel Size & 15um\x15um \\
  Image Area & 49.2mm\x49.2mm \\
  Cooling Mode & Cry-tiger: \~-$130^{\circ}$C \\
  Gain & 1.37e- \\
  Readout Noise & ~11.0e- \\
  \noalign{\smallskip}\hline
\end{tabular}
\ec
\end{table}

\begin{figure}
	\centering
	\includegraphics[width=12.0cm, angle=0]{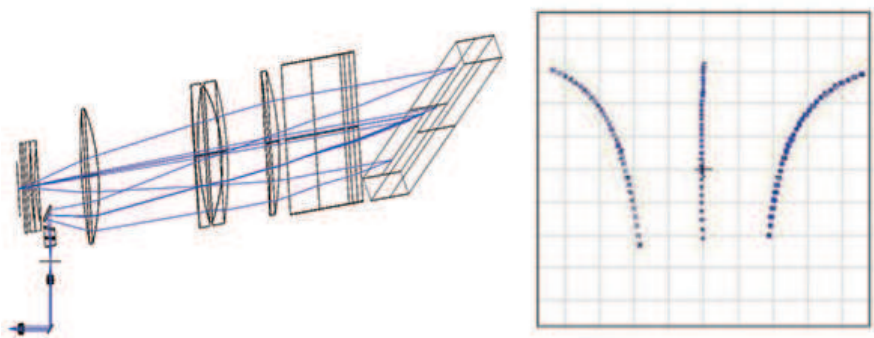}
	\caption{Layout of the DEM mode(left) and spectra format on the CCD(right)}
	\label{fig10}
\end{figure}

\begin{figure}
	\centering
	\includegraphics[width=12.0cm, angle=0]{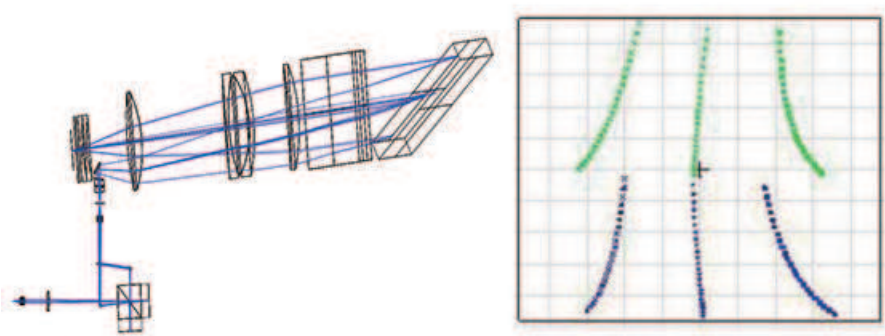}
	\caption{Layout of Radial Velocity mode(left) and spectra format on the CCD(right)}
	\label{fig11}
\end{figure}

\subsection{CHILI}
CHILI is based on a stock VIRUS unit developed for HETDEX. We can perform 2D spectral observations for extended sources with this instrument. The unit has two spectrograph channels with fixed angles between the collimator and camera. The fiber integral field unit(IFU) is mounted at its input. At the output, the fibers are arrayed into two linear arrays, each has 247 fibers. The IFU consists of 494 fibers with the core diameter of 200 microns. It is arranged in a hexagonal matrix at the input under a configuration with 23 rows of 11 or 12 fibers fed by micro-lens focal reducer optics. The field-of-view is 71" \x 76". It can be fully filled with the observational sky region. The micro-lens fill factor is at least 95\%, and it is not necessary to dither the telescope in order to fill in the observational sky region. The calibration source is Hg, Cd, and Ne emission line lamps. The optical diagram and the structure of CHILI is shown in figure 12, and some key parameters are described in table 8.

\begin{table}
\bc
\begin{minipage}[]{140mm}
\caption[]{Key Parameters Describing CHILI\label{tab8}}\end{minipage}
\setlength{\tabcolsep}{10pt}
\small
 \begin{tabular}{ccc}
  \hline\noalign{\smallskip}
  CCD & Number of Pixels & 2048\x2048 \\
   & Pixel Size & 15um\x15um \\
  Grating 1 & Resolving Power & 900(350nm-550nm) \\
  Grating 2 & Resolving Power & 900(460nm-720nm) \\
  \noalign{\smallskip}\hline
\end{tabular}
\ec
\end{table}

\begin{figure}
	\centering
	\includegraphics[width=14.0cm, angle=0]{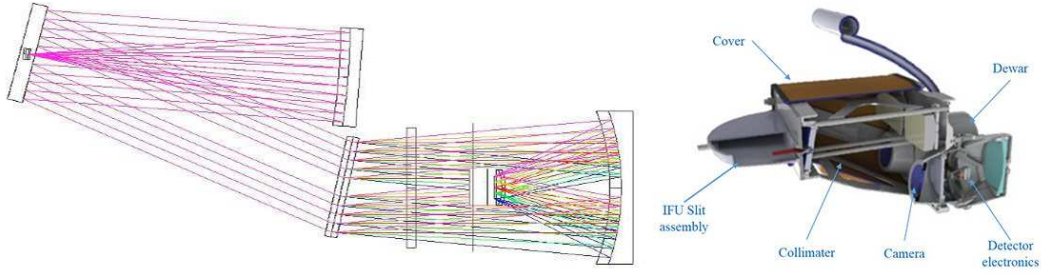}
	\caption{The Optical Diagram and the Structure of CHILI}
	\label{fig12}
\end{figure}

\subsection{Calibration}
We take dozens of the flux standard stars  at the photometric nights routinely and derive the average extinction curve of Lijiang observatory. Table 9 listed all the Landolt standard stars that were observed on Nov. 6 \& 7, 2013. The exposure time of each band is U=90s, B=30s, V=10s, R=10s, I=10s, respectively. All the CCD images were corrected for bias, flat field, and cleaned of cosmic rays, using IRAF. The results are shown in figure 13. It is made by the low-dispersion spectra (G3+Slit 10.0") and UBVRI filters with YFOSC. 

\begin{table}
\bc
\begin{minipage}[]{140mm}
\caption[]{Landolt Standard Stars Used in this Paper\label{tab9}}\end{minipage}
\setlength{\tabcolsep}{10pt}
\small
 \begin{tabular}{cccccccc}
  \hline\noalign{\smallskip}
  Star & $\alpha$(J2000) & $\delta$(J2000) & V & U-B & B-V & V-R & V-I \\
  SA92$\_$245 & 00:54:16 & +00:39:51 & 13.818 & 1.418 & 1.189 & 1.189 & 1.836 \\
  SA92$\_$248 & 00:54:31 & +00:40:15 & 15.346 & 1.128 & 1.289 & 1.289 & 1.243 \\
 SA92$\_$249 & 00:54:34 & +00:41:05 & 14.325 & 0.699 & 0.24 & 0.24 & 0.769 \\
 SA92$\_$250 & 00:54:37 & +00:38:56 & 13.178 & 0.814 & 0.48 & 0.48 & 0.840 \\
 SA92$\_$425 & 00:55:59 & +00:52:58 & 13.941 & 1.191 & 1.173 & 1.173 & 1.382 \\
 SA92$\_$426 & 00:56:00 & +00:52:53 & 14.466 & 0.729 & 0.184 & 0.184 & 0.808 \\
 SA92$\_$355 & 00:56:06 & +00:50:47 & 14.965 & 1.164 & 1.201 & 1.201 & 1.404 \\
 SA92$\_$430 & 00:56:16 & +00:53:16 & 14.440 & 0.567 & -0.04 & -0.04 & 0.676 \\
 SA95$\_$330 & 03:54:31 & +00:29:05 & 12.174 & 1.999 & 2.233 & 2.233 & 2.266 \\
 SA95$\_$275 & 03:54:44 & +00:27:20 & 13.479 & 1.763 & 1.74 & 1.74 & 1.942 \\
 SA95$\_$276 & 03:54:46 & +00:27:20 & 14.118 & 1.225 & 1.218 & 1.218 & 1.394 \\
 SA95$\_$112 & 03:53:40 & -00:01:13 & 15.502 & 0.662 & 0.077 & 0.077 & 1.225 \\
 SA95$\_$41 & 03:53:41 & -00:02:31 & 14.06 & 0.903 & 0.297 & 0.297 & 1.174 \\
 SA95$\_$42 & 03:53:44 & -00:04:33 & 15.606 & -0.215 & -1.111 & -1.111 & -0.299 \\
 SA95$\_$115 & 03:53:44 & -00:00:48 & 14.680 & 0.836 & 0.096 & 0.096 & 1.156 \\
 SA95$\_$43 & 03:53:49 & -00:03:01 & 10.803 & 0.51 & -0.016 & -0.016 & 0.624 \\
 SA92$\_$410 & 00:55:15 & +01:01:49 & 14.984 & 0.398 & -0.134 & -0.134 & 0.481 \\
 SA92$\_$412 & 00:55:16 & +01:01:53 & 15.036 & 0.457 & -0.152 & -0.152 & 0.589 \\
 SA95$\_$328 & 03:54:19 & +00:36:28 & 13.525 & 1.532 & 1.298 & 1.298 & 1.776 \\
 SA95$\_$329 & 03:54:24 & +00:37:07 & 14.617 & 1.184 & 1.093 & 1.093 & 1.408 \\
 
  \noalign{\smallskip}\hline
\end{tabular}
\ec
\end{table}

The observations of the same targets on Nov. 6 \& 7, 2013 can be used to obtain the signal-noise ratio(S/N) among UBVRI-bands in different photometric system. Figure 14 shows the comparison between YFOSC and PI CCD among UBVRI-bands for the same targets. The S/N and Magnitude in figure 14 are defined by the following equations: 

\begin{equation}
\frac{S}{N}=\frac{N_{star}}{\sqrt{N_{star}+n_{pix}(N_{sky}+N_{dark}+N_{readout}^2)}}
\end{equation}
\begin{equation}
Magnitude = M_{stand} + KX
\end{equation}

where $N_{star}$ is the photons collected from targets; $N_{sky}$ is the photons per pixel from the sky background; $N_{dark}$ is the CCD dark current per pixel; $N_{readout}$ is the readout noise; $n_{pix}$ is the FWHM of each target; $M_{stand}$ is the magnitude in Landolt photometry system for the targets; $K$ is the extinction coefficient and $X$ is the airmass. For YFOSC and PI CCD, the readout noise is well estimated and the dark current is negligible. Generally, YFOSC has higher S/N than PI CCD, especially in U and I bands.

 The same observational data can also be used to estimate the efficiency of the photometric system in UBVRI-bands. Table 10 gives the efficiencies from the telescope to the CCD of YFOSC and PI CCD at two epochs, respectively. The throughput estimated in January and November can imply the throughput decreasing of telescope. It also suggests the effect of the mirror re-coating. 
 
\begin{figure}
	\centering
	\includegraphics[width=9.0cm, angle=0]{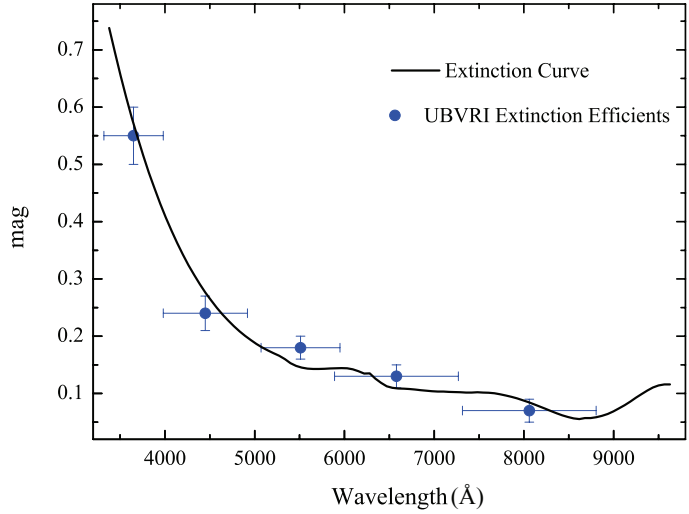}
	\caption{Average atmospheric extinction curve and coefficients at Lijiang observatory. The curve obtained by YFOSC long-slit spectrograph and over-plotted with coefficients of UBVRI-bands photometry. The horizontal error bars of coefficients stand for the FWHM of the filters and vertical error bars for the measurement error.}
	\label{fig13}
\end{figure}
\begin{figure}
	\centering
	\includegraphics[width=9.0cm, angle=0]{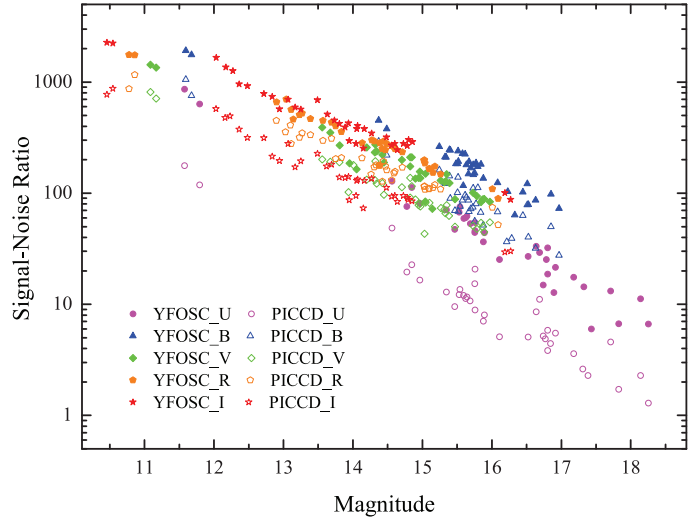}
	\caption{Signal noise ratio comparisons between YFOSC and PI CCD among UBVRI bands}
	\label{fig14}
\end{figure}
\begin{table}
\bc
\begin{minipage}[]{140mm}
\caption[]{Efficiency from telescope to CCD of PICCD and YFOSC in UBVRI\label{tab10}}\end{minipage}
\setlength{\tabcolsep}{10pt}
\small
 \begin{tabular}{cccc}
  \hline\noalign{\smallskip}
  Filter & PICCD(\%) & YFOSC(\%) & Ratio(YF/PI) \\
  \hline\noalign{\smallskip}
   & & Jan. 2013 & \\
  U & 3.9$\pm$0.2 & 8.9$\pm$0.5 & 2.27 \\
  B & 12.5$\pm$0.3 & 15.3$\pm$0.4 & 1.23 \\
  V & 22.4$\pm$0.4 & 22.5$\pm$0.3 & 1.01 \\
  R & 21.2$\pm$0.3 & 20.3$\pm$0.4 & 0.96 \\
  I & 17.1$\pm$0.4 & 34.5$\pm$0.5 & 2.02 \\
  \hline\noalign{\smallskip}
  \hline\noalign{\smallskip}
   & & Nov. 2013 & \\
  U & 3.2$\pm$0.3 & 7.3$\pm$0.4 & 2.29 \\
  B & 10.0$\pm$0.4 & 12.4$\pm$0.4 & 1.24 \\
  V & 18.3$\pm$0.3 & 18.0$\pm$0.3 & 0.98 \\
  R & 17.2$\pm$0.3 & 16.3$\pm$0.4 & 0.96 \\
  I & 13.8$\pm$0.4 & 28.1$\pm$0.4 & 2.03 \\
  \noalign{\smallskip}\hline
\end{tabular}
\ec
\end{table}

\section{Auxiliary Equipment}
\label{sect:Equipment}
\subsection{Dome}
	The dome of the 2.4-meter telescope is a classical type with the shuttles that can open and close during observations. The main purpose for using this type of dome is to prevent the influence from wind. In order to improve the dome seeing, there have eight side windows and four air channels used to balance the temperature inside and outside the dome. There has an independent PLC system that can communicate with the MCC and receive the control request from the TCC to open/close the shuttle and to track the target along with the telescope.
	
	The dome is a super-hemisphere sphere with a steel welded spherical grid structure. The dome mainly consists of 8 parts, including chassis, arched girder, beams, shuttle, shuttle driver, azimuth driver, electric control system and a crane inside. The dome skeleton is formed by welded angle steel, and the arched girder is welded by the 12mm steel plate. The steel channel can guarantee the stiffness requirement of the dome and can also be used as the foundation support for the crane. The open range of the shuttle is from -$3^{\circ}$ to $100^{\circ}$by using the fixed chain driver. It has dual protection by using the electrical-limit switch and the mechanical-limit instruments. The chassis of the dome uses wheel rail structure. It is driven by the electrical motors with a decelerating gear, and it can move the dome in $360^{\circ}$ along the azimuth.

\subsection{Coating Machine}
	The coating machine(ZZS3200) is developed by Yunnan Observatories and the Chengdu Institute of Optical and Electronic Technology. It is used for the 2.4-meter telescope and 1.8-meter telescope in Lijiang Observatory. The inner diameter of the machine is 3200mm, and the height of the machine is 3500mm. It is divided into two parts. The lower part is mounted on the rail that can move out to put a mirror on it. We re-coated the primary mirror of the 2.4-meter telescope  using this coating machine in October 2012, and the optical efficiency is increased significantly. The comparison of the same target before and after the re-coating is shown in table 11. We can re-coat the primary mirror of the 2.4-meter telescope every 2 years in order to keep the excellent optical efficiency.
	
\begin{table}
\bc
\begin{minipage}[]{140mm}
\caption[]{Comparison of same target before and after re-coating\label{tab11}}\end{minipage}
\setlength{\tabcolsep}{10pt}
\small
 \begin{tabular}{ccc}
  \hline\noalign{\smallskip}
   & Before(ADU) & After(ADU) \\
  Observing & 2011-12-20 & 2012-12-13 \\
  Time & BT19:32:30 & BT 19:53:33 \\
  PG2331$+$055 & 173888 & 355453 \\
  PG2331$+$055A & 1203000 & 2377830 \\
  PG2331$+$055B & 259466 & 503664 \\
  \noalign{\smallskip}\hline
\end{tabular}
\ec
\end{table}

\subsection{Astronomical Site Monitoring System}
	Astronomical Site Monitoring System(ASMS) is a system that monitoring the weather condition, astronomical condition, and instruments of the site. The system can provide the information like weather, seeing condition, cloudy, all-sky image and so on. Besides, it can also provide the weather data to the TCS for protecting the telescope and can add the weather information to the head of the observation data files to help the data processing. This system consists of weather station and monitoring system. It can be accessed through the internet\footnote {http://weather.gmg.org.cn:9000}.

\section{Scientific Achievement}
\label{sect:scientific}
	Many scientific results have been obtained from the Lijiang 2.4-meter telescope. For example, with the low-dispersion spectrograph of YFOSC,\cite{2015Natur.518..512W}  found the highest luminosity quasar in the early universe (z$\sim$6.3). The ultra-high luminosity of this quasar indicates that there may have a 12 billion solar-mass black hole in its core. This black hole is a magnitude bigger than the others found before. How can a black hole with such mass be formed in less than 1 billion years after the birth of the universe? This discovery raises new questions about the evolution of black holes and even the evolution of the universe.

By using the YFOSC, \cite{2014ApJ...793..108W}, \cite{2014ApJ...782...45D, 2015ApJ...806...22D, 2016ApJ...820...27D,2016ApJ...825..126D},\cite{2016ApJ...827..118L},and \cite{2016ApJ...822....4L} performed reverberation-mapping observations of some high accretion rate AGNs. They have measured the masses of a set of high accretion rate black holes, in order to study the high-redshift cosmology.

Hundreds of observations on the supernova have been performed with this telescope in the past five years based on the LiONS (Li-Jiang One hour per Night observation of Supernovae) project. This project focuses on the type Ia supernovae at a very early phase or any peculiar events of supernovae, and can monitor the interesting targets in low-resolution spectroscopy and multi-band photometry immediately and frequently(e.g., \citealt{2017Natur.551..210A} ; \citealt{2015ApJ...807...59H,2016ApJ...832..139H}; \citealt{2018MNRAS.478.4575L}; \citealt{2015MNRAS.452..838L}; \citealt{2016AJ....151..125Z} ; \citealt{2014AJ....148....1Z, 2014ApJ...797....5Z, 2016ApJ...817..114Z, 2018ApJ...863..109Z}).

Some studies focus on the intergalactic objects. \cite{2015ApJ...798L..42Q}  carried out optical variation study of the binary star using 2.4-meter telescope. They found the first stable red dwarf binary. \cite{2011MNRAS.414L..16Q}  discovered one exoplanet that is surrounding to an evolutionary binary star. \cite{2014MNRAS.437.2566S}  measured some stellar parameters for the study of the white dwarfs.

Gamma-ray burst follow-up observation is also performed by Lijiang 2.4-meter telescope. \cite{2012A&A...538A...1M}  studied the GRB100219A optical afterglow. We also perform the observation for the electromagnetic counterparts of the gravitational waves cooperated with world-wide astronomers.

\section{Summary and Future Expectation}
\label{sect:future}
The Lijiang 2.4-meter telescope is the largest optical telescope at present in China for general astronomical observation. After 10-years operation, it has equipped many kinds of observational instruments covering photometry, low/medium resolution spectral, high-resolution spectral, and two-dimensional spectral observation. Many kinds of scientific research can be performed by this telescope, and a set of scientific achievements have been obtained. More than 190 papers related to the Lijiang 2.4-meter telescope have been published.

Apart from the instruments described above, there are two new instruments under construct. One is the EMCCD used for both high precision temporal observation and high-space resolution photometry with lucky imaging techniques. The other is the near-infrared spectrograph(ONICE) used for the infrared spectral observation. Besides, observers can also install their own specified instruments onto Cassegrain side port using the interface plate provided. With these new instruments, the Lijiang 2.4-meter telescope can be involved in more and more scientific research topics. It will play a more unique role in the observational astronomy in the future.

\normalem
\begin{acknowledgements}
We thank all the present and former staffs at Lijiang Observatory for their effort on installation, operating and maintaining the Lijiang 2.4-meter telescope, and the staffs who help to develop the instruments. This work is supported by the Joint Research Fund in Astronomy (U1631127, U1631129, U1831204) under cooperative agreement between the National Natural Science Foundation of China (NSFC) and Chinese Academy of Science (CAS), the National Natural Science Foundation of China (NSFC) (11473068,11603072,11573067), the National Key R\&D Program of China (2018YFA0404603),  and also supported by the Key Laboratory for the Structure and Evolution of Celestial Objects,Chinese Academy of Sciences (CAS).

\end{acknowledgements}


\end{document}